\documentstyle[preprint,aps,epsf]{revtex}

\catcode`\@=11
\def\lsim{\mathrel{\mathpalette\@versim<}}
\def\gsim{\mathrel{\mathpalette\@versim>}}
\def\@versim#1#2{\vcenter{\offinterlineskip
\ialign{$\m@th#1\hfil##\hfil$\crcr#2\crcr\sim\crcr } }}
\catcode`\@=12

\begin{document}

\tightenlines

\draft
\preprint{KANAZAWA-02-12}
\title{Suppressing FCNC and CP-Violating Phases\\
by Extra Dimensions}
\author{Jisuke Kubo and Haruhiko Terao}
\address{
Institute for Theoretical Physics, 
Kanazawa  University, 
Kanazawa 920-1192, Japan
}
\maketitle
\begin{abstract}
In extra dimensions the infrared attractive force of gauge interactions is
amplified. We find that this force can align in the infrared limit the 
soft-supersymmetry breaking terms out of their anarchical disorder at a 
fundamental scale, in such a way that flavor-changing neutral currents as well
as dangerous CP-violating phases are sufficiently suppressed at the unification
scale. The main assumption is that the matter and Higgs supermultiplets and the
flavor-dependent interactions such as Yukawa interactions are stuck at the 
four-dimensional boundary. As a concrete example we consider the minimal model
based on $SU(5)$ in six dimensions.
\end{abstract}



Low-energy softly-broken supersymmetry (SUSY) has 
been the most promising idea in solving
the gauge hierarchy problem \cite{susy}. However, the 
introduction of the superpartners of the known particles
induces
large flavor-changing neutral current (FCNC) processes and 
CP-violating phases, which are 
severely constrained by precision experiments  \cite{fcnc}. 
Therefore, the huge degrees of
freedom involved in  
the soft-supersymmetry breaking (SSB) parameters
have to be highly constrained in all viable 
supersymmetric models. 
This has been called the supersymmetric flavor problem.

To overcome  this problem, several ideas
of SUSY breaking and its mediation mechanisms 
have been proposed; gauge mediation \cite{gauge},
anomaly mediation \cite{anomaly}, gaugino mediation \cite{gaugino} 
and so on.
The common feature behind these ideas is that the SSB
parameters are generated by flavor-blind radiative corrections,
and that their tree-level contributions at a fundamental scale
$M_{\rm PL}$ 
are assumed to be sufficiently
suppressed, {\it e.g.}, by sequestering of branes for the 
visible sector and the hidden SSB sector. 
However,  it has been  argued recently \cite{string} 
that such a sequestering mechanism cannot be simply 
realized in generic supergravity or superstring inspired models.
Therefore,
the supersymmetric flavor problem 
still manifests itself in
the above-mentioned mediation mechanisms and their modifications.

An interesting way out from this problem is to suppress the tree-level
contributions by certain field theoretical dynamics.
There have been indeed  several attempts along the line of
thought,  in which use has been made \cite{ns,nkt,ls}
that the SSB parameters
are suppressed in the infrared limit in
approximate superconformal field theories \cite{karch}.
In this letter, we propose another possibility
in more than four dimensions that
flavor-blind radiative corrections 
dominate over the tree-level contributions
as well as  any other flavor non-universal corrections.

Let us start to present our idea.
We assume that only the supersymmetric 
gauge interactions
exist  in the $(4+\delta)$ dimensional bulk 
($\delta =1~\mbox{or}~2$),  while 
all the other interactions are confined 
at the four-dimensional 
boundary \cite{antoniadis1,dienes1,arkani1}.
Accordingly, the $(4+\delta)$-dimensional
gauge supermultiplet
propagates in the bulk, and  all the 
$N=1$ chiral supermultiplets $\Phi_i =(\phi_i, \psi_i)$
containing  matters and
Higgses propagate only in four dimensions.
The gauge supermultiplet contains 
a chiral supermultiplet $ \Gamma $ in the adjoint
representation, to which
we assign an odd parity \cite{dienes1,arkani1}
so that it does not contain zero modes, and does not have any interactions
with $\Phi$'s.
To simplify the situation, we further assume that
each extra dimension is compactified on
a circle with the same radius $R$.
The size of $R$ is model dependent, but
throughout this letter we
assume that $M_{\rm GUT}=1/R$.
The boundary superpotential
has a generic form
\begin{equation}
W(\Phi) = \frac{1}{6} Y^{ijk} \Phi_i \Phi_j \Phi_k + 
\frac{1}{2} \mu^{ij} \Phi_i\Phi_j ~,
\end{equation}
and that the SSB Lagrangian $L_{\rm SSB}$ can be written as
\begin{equation}
-L_{\rm SSB}= \left(\frac{1}{6} 
 h^{ijk} \phi_i \phi_j \phi_k 
 +  \frac{1}{2}  B^{ij} \phi_i \phi_j 
+  \frac{1}{2} \sum_{n=0} M\lambda_n \lambda_n+
\mbox{h.c.}\right)
+\phi^{*j}(m^2)^i_j \phi_j~,
\label{ssbL}
\end{equation}
where $\lambda_n$'s are the Kaluza-Klein modes
of the gaugino, and 
we have assumed  a unique gaugino mass $M$
for all $\lambda$'s.

We  consider the renormalization group (RG) running of the parameters
between the fundamental scale
$M_{\rm PL}=M_{\rm Planck}/\sqrt{8\pi}\simeq
2.4 \times 10^{18}$ GeV and 
$M_{\rm GUT}$.
To see the gross behavior of
the RG running, we consider
the contributions coming
from only the gauge supermultiplet,
because it is  the only
source responsible for the power-law running \cite{veneziano,dienes1}
of the parameters under the assumptions specified above.
We find the following set of the  one-loop  $\beta$ functions
in this approximation \cite{dienes1,kobayashi1}:
\begin{eqnarray} 
\Lambda \frac{d g }{d\Lambda} &=&
-\frac{2}{16\pi^2}C(G)G_\delta^2g,\\
\Lambda \frac{d M }{d\Lambda} &=&
-\frac{4}{16\pi^2}  C(G)G_\delta^2 M,
\label{betag1}\\
\Lambda \frac{d Y^{ijk} }{d\Lambda}&=&
-\frac{2}{16\pi^2} (C(i)+C(j)+C(k))G_\delta^2 Y^{ijk},
\label{betaY1}\\
\Lambda \frac{d \mu^{ij} }{d\Lambda} &=&
-\frac{2}{16\pi^2} (C(i)+C(j))G_\delta^2~\mu^{ij},
\label{betamu1}\\
\Lambda \frac{d B^{ij} }{d\Lambda} &=&
\frac{2}{16\pi^2}(C(i)+C(j))G_\delta^2
(2 M\mu^{ij}-B^{ij}),
\label{betab1}\\
\Lambda \frac{d h^{ijk} }{d\Lambda} &=&
\frac{2}{16\pi^2} (C(i)+C(j)+C(k))G_\delta^2 (2 M Y^{ijk}-h^{ijk}),
\label{betah1}\\
\Lambda \frac{ d (m^2)^i_j }{ d\Lambda } &=&
- \frac{8}{16\pi^2} C(i)\delta_j^i G_\delta^2|M|^2,
\label{betam21}
\end{eqnarray}
where
$G_\delta = g X_\delta^{1/2}(R\Lambda)^{\delta/2}$, and
$X_\delta=\pi^{\delta/2} \Gamma^{-1}(1+\delta/2)
=2 (\pi)$ for $\delta=1 (2)$ \cite{dienes1}.
The gauge coupling is denoted by $g$, and 
$C(G)$ stands for the
quadratic Casimir of the adjoint representation of the
gauge group $G$, and $C(i)$
for  that of the representation $R_{i}$.
It is easy to show that the evolution of $Y^{ijk}, \mu^{ij}$ 
and $M$ are related to that of $g$ as
\begin{eqnarray}
M(M_{\rm GUT}) &=& 
\left(\frac{g(M_{\rm GUT})}{g(M_{\rm PL})}\right)^{2}
M(M_{\rm PL}), \\
Y^{ijk}(M_{\rm GUT}) &=& 
\left(\frac{g(M_{\rm GUT})}{g(M_{\rm PL})}\right)^{\eta_Y^{ijk}}
Y^{ijk}(M_{\rm PL}),
\label{yijk}\\
\mu^{ij}(M_{\rm GUT}) &=&
\left(\frac{g(M_{\rm GUT})}{g(M_{\rm PL})}\right)^{\eta_\mu^{ij}}
\mu^{ij}(M_{\rm PL}),
\label{muij}
\end{eqnarray}
where the exponents are given as 
$\eta_Y^{ijk}=(C(i)+C(j)+C(k))/C(G)$ and 
$\eta_\mu^{ij}=(C(i)+C(j))/C(G)$.
Therefore these parameters can become very large
if $g(M_{\rm PL})/g(M_{\rm GUT})$ is large. 
A rough  estimate shows that
\begin{equation}
\frac{g(M_{\rm GUT})}{g(M_{\rm PL})} \simeq 
\left[ \frac{ C(G)X_{\delta} \alpha_{\rm GUT}}{\pi \delta}
\right]^{1/2}
\left(\frac{M_{\rm PL}}{M_{\rm GUT}}\right)^{\delta/2}.
\label{ratio-g}
\end{equation}
If we use 
$\alpha_{\rm GUT}=0.04, M_{\rm PL}/M_{\rm GUT}=10^2, G=SU(5)$,
then this is given approximately by
3.5 for $\delta=1$ and 32 for $\delta=2$.

The ratios between other SSB  parameters to the gaugino mass $M$
approach to their infrared attractive fixed points;
\begin{eqnarray}
B^{ij}/M &\to & -\eta_\mu^{ij} \mu^{ij}, \nonumber \\
h^{ijk}/M & \to & -\eta_Y^{ijk} Y^{ijk},  \nonumber \\
(m^2)^{i}_{j}/|M|^2 &\to & \left(C(i)/C(G)\right)\delta_j^i,
\label{ssb4}
\end{eqnarray}
where $\eta$'s are used in (\ref{yijk}) and (\ref{muij}).
Note that so far no assumption on the reality of the 
parameters has been made. 
We see that
the low-energy structure is 
completely fixed by the group theoretic structure of the model.
Furthermore, since $h^{ijk}$ and $(m^2)^{i}_{j}$ 
become aligned in the infrared limit, 
{\it i.e.}, $h^{ijk} \propto Y^{ijk}$ and $(m^2)^{i}_{j} \propto
\delta_j^i$, the infrared forms (\ref{ssb4})
give desired initial values of the parameters at $M_{\rm GUT}$ 
to suppress FCNC processes  in the MSSM, and they predict that the 
only CP-violating phase is the usual CKM phase.

One can easily estimate how much of
a disorder in the initial values at $M_{\rm PL}$
can survive at $M_{\rm GUT}$.
Suppose that there exists an $O(1)$ disorder in $(m^2)^i_j/|M|^2$. 
Using the $\beta$ functions 
(\ref{betag1}) and (\ref{betam21}), we find the deviation from
(\ref{ssb4}) to be
\begin{equation}
\left(\frac{g(M_{\rm PL} )}{g(M_{\rm GUT} )}\right)^4
\left[\frac{(m^2)^i_j}{|M|^2}(M_{\rm PL})  
-\frac{C(i)}{C(G)}\delta^i_j\right].
\label{disorderm}
\end{equation}
Then inserting the value of $ g(M_{\rm PL} )/g(M_{\rm GUT})$
given in (\ref{ratio-g}), 
we find that 
an $O(1)$ disorder at $M_{\rm PL}$ becomes
a disorder of $O(10^{-2})$ and $O(10^{-6})$  at $M_{\rm GUT}$
for $\delta=1$ and $2$,
respectively. Note that the off-diagonal elements of $(m^2)^i_j$ 
as well as the differences among the diagonal elements
$\Delta m^2(i,j)=(m^2)^i_i-(m^2)^j_j$ (if $C(i)=C(j)$)
belong to the disorder. However, their contributions to
$(\delta_{ij})_{LL,RR}$ of \cite{fcnc} are less than 
$O(10^{-6})$ for $\delta=2$, therefore
the most stringent constraint coming from
$\mu\to e\gamma$ is satisfied \cite{fcnc}.
However, in the case of five dimensions ($\delta=1$) the
suppression of the disorder will not be sufficient.
[If we use $M_{\rm PL}/M_{\rm GUT} \sim 10^3$, then the suppression
is much improved and the five dimensional case is also allowed.]

Similarly, using (\ref{betag1}) and (\ref{betah1}), we obtain
the deviation for the tri-linear couplings from (\ref{ssb4}) as
\begin{eqnarray}
\left(\frac{g(M_{\rm PL} )}{g(M_{\rm GUT} )}\right)^2
\left[\frac{h^{ijk}}{M}(M_{\rm PL})  
+\eta^{ijk}_Y Y^{ijk} (M_{\rm PL})\right],
\label{disorderh}
\end{eqnarray}
where use has been made of (\ref{yijk}).
Suppose the tri-linear couplings to be order of $M Y^{ijk}$
at $M_{\rm PL}$.
Then we find that 
\begin{equation}
\left|
\frac{h^{ijk}}{\eta_Y^{ijk}M Y^{ijk}}(M_{\rm GUT}) + 1
\right|
\lsim 
\left(\frac{g(M_{\rm PL})}{g(M_{\rm GUT})}\right)^{2+\eta_Y^{ijk}}.
\label{disorderh2}
\end{equation}
Note that the phases of $h^{ijk}/M Y^{ijk}$ can be suppressed.
In the case of $G=SU(5)$, $\eta_Y^{ijk}=48/25 (42/25)$
for the up (down) type Yukawa couplings.
Using  (\ref{ratio-g}) again, we find that the right-hand side of
(\ref{disorderh2}) is
$\sim 10^{-2(6)}$ for $\delta=1 (2)$.
This disorder contributes, for instance,
to $Im (\delta_{ii})_{LR}$ 
as well as $Re(\delta_{ij})_{LR}$ of \cite{fcnc}.
Therefore  our  suppression  mechanism  can satisfy
the most stringent constraints coming from
the electric dipole moments (EDM) of the neutron and the 
electron \cite{fcnc}.
Similarly the phases of the B-parameter, $B^{ij}/M \mu^{ij}$ 
are also suppressed.

In concrete examples, there will be logarithmic corrections
to (\ref{ssb4}) to which the Yukawa couplings $Y^{ijk}$
non-trivially contribute.
How much 
the logarithmic corrections can
amplify the disorder
will  be model-dependent.
It is certainly worthwhile to note that
 the logarithmic interactions
will be non-negligible only  for $\Lambda$ close to $M_{\rm GUT}$,
thereby overcoming the problem found in \cite{polonsky}
that the GUT effects may destroy
the universality of the SSB terms.

To be more specific, we consider the 
minimal GUT model based on $G=SU(5)$ in six dimensions.
The boundary fields are: chiral superfields 
$\Psi^{i}({\bf 10})$ and $\Phi^{i}(\overline{\bf 5})$,
where $i$ runs over the three generations, 
$\Sigma({\bf 24})$ to break $SU(5)$,
and two Higgs superfields
$H({\bf 5})$ and $\overline{H}({\overline{\bf 5}})$.
To simplify the situation, we neglect the neutrino masses and their mixings. 
The boundary superpotential of the model is given by
\begin{equation}
W = \frac{Y_{U}^{ij}}{4}\,
\Psi^{i}\Psi^{j}H+
\sqrt{2}\,Y_D^{ij}\,\Phi^{i}
\Psi^{j}\overline{H}+
\frac{Y_{\lambda}}{3}\,\Sigma^3 
+Y_{f}\,\overline{H}\Sigma H+ \frac{\mu_{\Sigma}}{2}\,
\Sigma^2+ \mu_{H}\,\overline{H} H, 
\end{equation}
where $Y_{U}^{ij}$ and $Y_{D}^{ij}$
are the Yukawa couplings.
Then the gross infrared attractive
form of the SSB parameters (\ref{ssb4}) becomes
\begin{eqnarray}
& &
B_{\Sigma} \to -2 M \mu_{\Sigma},~
B_{H}\to -\frac{24}{25} M \mu_{H}, 
\label{infra-B}\\
& &
h_{U}^{ij} \to -\frac{48}{25} M Y_{U}^{ij},~
h_{D}^{ij} \to -\frac{42}{25} M Y_{D}^{ij}, 
\label{infra-h}\\
& &
(m^2_{\Phi})^{ij} \to \frac{12}{25} |M|^2 \delta^{ij},~
(m^2_{\Psi})^{ij} \to \frac{18}{25} |M|^2\delta^{ij},
\label{infra-m}\\
& &
m^2_{H_d}, m^2_{H_u} \to \frac{12}{25} |M|^2,
\end{eqnarray}
in an obvious notation.
We find also
$h_{f} \to -(49/25) M Y_{f},~
h_{\lambda} \to -3M Y_{\lambda},~
m^2_{\Sigma} \to 2~|M|^2$.
All the scalars that belong to ${\bf 5}$ or $\overline{{\bf 5}}$
have the same positive squared soft-mass ($\approx (0.69 M)^2$),
which does not differ very much from
that ($\approx (0.85 M)^2$) for the scalars belonging
to ${\bf 10}$. 
So, the infrared attractive form in the present model
is similar to the SSB terms of the constrained MSSM (CMSSM),
and therefore,
the model predicts a similar spectrum as in the CMSSM.

In the following analyses we would like to neglect 
the mixings of the matter multiplets, because
their effects will be very small as seen later.
One of the pleasant feature of the infrared
attractive form of the SSB terms (\ref{ssb4})
is that the tri-linear couplings, too,  may be assumed to
be small if the corresponding Yukawa couplings 
are small, as we have seen in (\ref{disorderh}).
Consequently, we will approximate
$Y_{U,D}^{ij}$  by $Y_{t,b} \delta^{i3}\delta^{j3}$ and 
$h_{U,D}^{ij} = h_{t,b} \delta^{i3}\delta^{j3}$, respectively.

To proceed, we  require that  
the MSSM is the effective theory below $M_{\rm GUT}$,
and we would like to check whether the initial values of the SSB
terms at $M_{\rm GUT}$ given in (\ref{infra-B})-(\ref{infra-m})
cause conflicts at low-energies,
especially with electroweak symmetry breaking.
For simplicity, we assume that
the potential of the MSSM 
at $M_{\rm SUSY}$ takes the tree-level form, 
where  we identify $M_{\rm SUSY}$ with
the unified gaugino mass $M$.
We require that $M_t=174$ GeV and
$M_{\tau}=1.77$ GeV, while  imposing
the $b-\tau$ unification at $M_{\rm GUT}$.
[But we will not take the mass of the bottom quark very seriously.
 It becomes  slightly larger than its experimental value.]

We are also interested in how much  the Yukawa
interactions modify the infrared attractive
values (\ref{infra-B})--(\ref{infra-m}).
Because of the limitation of space, we present below only one
case, which is consistent with  electroweak symmetry breaking.
The parameters of the case are:
$M=500~\mbox{GeV}~,~g = (0.0406\times 4\pi)^{1/2}~,~
M_{\rm GUT} =1.83 \times 10^{16}~\mbox{GeV}$, and
\begin{equation}
\mu_H = 935~\mbox{GeV},~
Y_t= 0.767 g,~
Y_b= 0.201 g,~
Y_f= 1.0 g,~
Y_\lambda =0.01 g,
\label{case2}
\end{equation}
where $\tan\beta$ is found to be $19.5$.
In this case the infrared attractive values of the SSB
terms are given by
\begin{eqnarray}
& &
(m^2_{\Phi^{1,2}}, m^2_{\Phi^{3}})
= (0.534, 0.531)|M|^2,~~
(m^2_{\Psi^{1,2}}, m^2_{\Psi^{3}})
= (0.801, 0.759)|M|^2, \nonumber \\
& &
(m^2_{H_u} , m^2_{H_d})
= (0.383, 0.420)|M|^2, \nonumber \\
& &
h_t = -1.97 M Y_t,~~
h_b = -1.74 M Y_b,~~
B_H = -0.922 M \mu_H.
\label{table1}
\end{eqnarray}
We have used the notation
$m_{\Phi^{i}}^{2}=(m_{\Phi}^{2})^{ii}$,
and similarly for $m_{\Psi^{i}}^{2}$.
It should be noted also that any charged sparticle
does not become a LSP with these soft scalar masses
at $M_{\rm GUT}$.

The $\beta$ functions for
$m_{\Phi^{1,2}}^{2}$ and  $m_{\Psi^{1,2}}^{2}$
do not depend on 
$Y_i$  and $h_i~(i=t,b,f,\lambda)$ in our 
approximation. Therefore, the infrared 
attractive values (\ref{infra-m})
are not modified by them.
There exist of course  logarithmic corrections coming
from the gauge interaction, but they are flavor-blind.
This is very pleasant, because the most stringent
constraint from FCNC processes
is the almost degeneracy of the squared soft masses
of the first two generations. 
We have found that for the initial values of 
$Y$'s and $g$ given in (\ref{case2}), 
the off-diagonal components $(m_{\Phi}^2)^{ij}/|M|^2$ 
and the difference of diagonal elements,
$\Delta m^2_{\Phi} (1,2)/|M|^2
=|m^2_{\Phi^{1}}-m^2_{\Phi^{2}}|/|M|^2$ 
(and  similarly for $m^2_{\Psi}$)
are less than $O(10^{-4})$,
which has been estimated to be $O(10^{-6})$
without the logarithmic corrections in (\ref{disorderm}).
This order of disorder at $M_{\rm GUT}$
is still sufficient to satisfy the
most stringent constraint 
coming from $\mu \to e\gamma$ \cite{fcnc}.
In contrast to the case of the first two generations,
the $\beta$ functions for 
$m_{\Phi^{3}}^{2}$ and $m_{\Psi^{3}}^{2}$ depend on 
$Y_i$  and $h_i$. Therefore, they  change their infrared 
attractive values.
Fig.~1  shows the evolution
of $m_{\Phi}^{2}/|M|^2$ and 
$m_{\Psi}^{2}/|M|^2$, respectively.
The dashed lines correspond to
the third generation.
We find that the differences 
$\Delta m^2_{\Psi} (i,3)/|M|^2=
|m_{\Psi^{i}}^{2}-m_{\Psi^{3}}^{2}|/|M|^2$
with $i=1,2$ 
at $M_{\rm GUT}$  are $\lsim 0.04$,  and 
they contribute to $\mu\to e\gamma$
through the mixing of the first two  and the third 
generations in the lepton sector,
{\it i.e.}, $V^{e *}_{3 i}V^{e}_{3 j}$ \cite{fcnc}.
Assuming that $V^{e}$ can be approximated by
the usual Cabibbo-Kobayashi-Maskawa matrix $V_{CKM}$,
we find that $(\delta_{12}^{\l})_{RR} \lsim 2\times 10^{-5}$
in the present case, so that 
$\Delta m^2_{\Psi} (i,3)/|M|^2$  does not 
cause a problem with
$\mu \to e \gamma$ \cite{fcnc}.
Similarly, the universality between $m_{H_{d}}^{2}$ and 
$m_{H_{u}}^{2}$ are also destroyed, as we see in (\ref{table1}).
The main origin are the top Yukawa coupling $Y_t$ and
$Y_f$. This does not conflict with the
FCNC problems and CP-violating processes.

In Fig.~2 the converging behavior for 
$-h_t/M Y_t$ and $-h_b/M Y_b$ are presented.
There is also no universality between $h_U$ and $h_D$ from
the beginning.
In (\ref{disorderh}) we have found that
the non-aligned part of $h^{ijk}$ is suppressed by a factor
of $10^{-6}$ in six dimensions, if the Yukawa couplings are neglected.
Let us estimate how much of this suppression can survive
if $Y$'s are taken into account. We find that
the corrections can be written as
$\Delta h_U^{ij}/M \sim 
(1/16\pi^2)
(~a_t Y_U^{i3} Y_U^{3j} Y_t +a_b Y_U^{i3} Y_D^{3j} Y_b)~
\ln \Lambda_{\rm eff}/M_{\rm GUT}$ 
and similarly for $\Delta h_D^{ij}/M $,
where $a_t$ and $a_b$
are $O(1)$ constants, and we have assumed that $h_{U,D}$ are proportional
to $M Y_{U,D}$ at a scale $\Lambda_{\rm eff}$, at which
$Y$'s become non-negligible. 
Further considerations 
in the basis where $Y_U$ is diagonal
yield that nonzero 
contributions (that are relevant to us) are:
$|\Delta h_U^{3j}(j\neq 3) |\sim  Y_t Y_b^2 L,~
|\Delta h_D^{i3} (i\neq 3)| \sim V_{CKM}^{i b} Y_t^2 Y_b L,~
|\Delta h_D^{ij}(i,j\neq 3)| \sim V_{CKM}^{i b}  Y_b^3 L$,
where $L=M \ln (\Lambda_{\rm eff}/M_{\rm GUT})/16\pi^2$.
Assuming that 
$\Lambda_{\rm eff} \sim 50 M_{\rm GUT}$, we find
that $ |\Delta h_U^{3j}(j\neq 3)/M |\sim O(10^{-4})$ for the values
given in (\ref{case2}), and the other $\Delta h$'s receive
a further suppression
from $V_{CKM}$.
$Im \Delta h_D^{11}$, for instance, contributes to
the EDM of the neutron, and can 
 be estimated to be  $O(10^{-7})$.
Therefore, we may conclude that the disorder of the trilinear couplings 
caused by the  Yukawa couplings
are sufficiently suppressed
to satisfy even the most
stringent constraints from the EDMs \cite{fcnc}.

We conclude that gauge interactions in extra dimensions
can be used to suppress 
the disorder of the SSB terms at the fundamental scale
so that the FCNC processes and dangerous 
CP-violating phases become tiny
at lower energy scales.
Thus the smallness of FCNC may be a possible hint of the 
existence of extra dimensions.

\begin{figure}[htb]
\epsfxsize=0.6 \textwidth
\centerline{\epsfbox{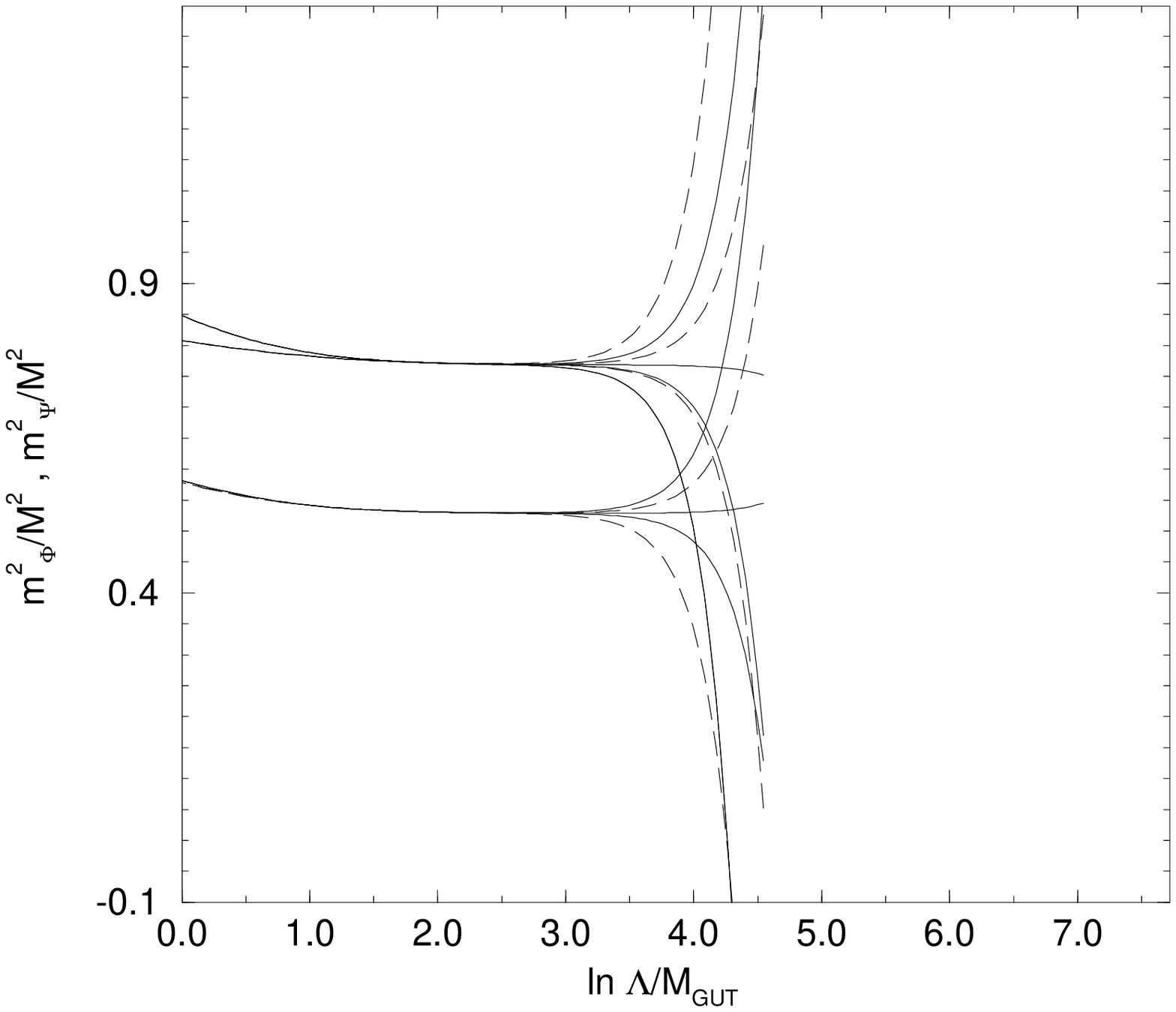}}
\caption{Infrared attractiveness of  $m^2_{\Phi}/|M|^2$ and
$m^2_{\Psi}/|M|^2$. The dashed (solid)  lines correspond to 
the third (first two) generation(s). 
$m^2_{\Psi^{1,2}} > m^2_{\Psi^{3}}
> m^2_{\Phi^{1,2}} \simeq m^2_{\Phi^{3}}$
at $\Lambda=M_{\rm GUT}$.}
\label{fig1}
\epsfxsize=0.6 \textwidth
\centerline{\epsfbox{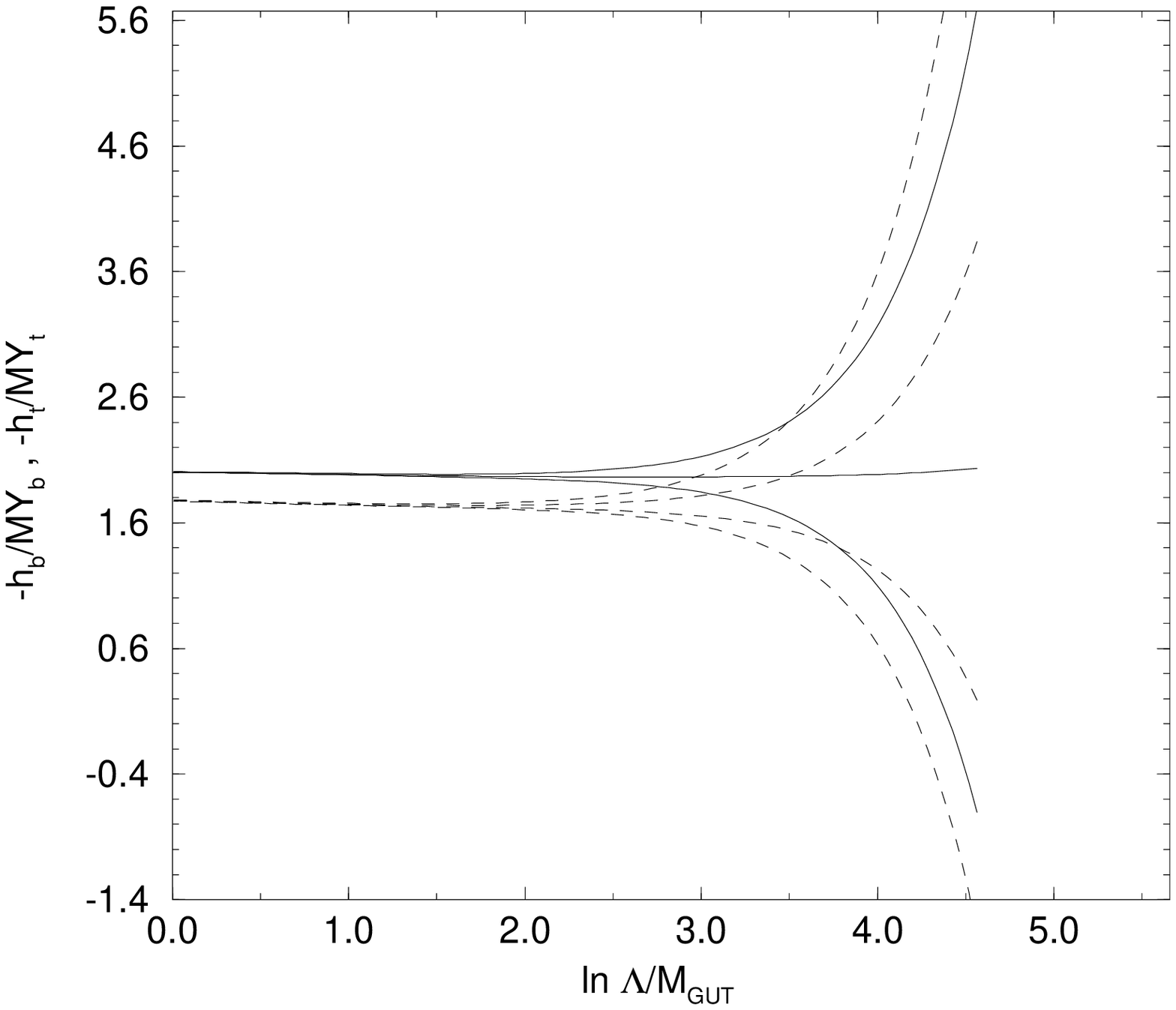}}
\caption{Infrared attractiveness 
of  $-h_{t}/M $ (solid) and $-h_{b}/M $ (dashed).}
\label{fig2}
\end{figure}


\begin{references}

\bibitem{susy}    
For a recent review, see, for instance,
G.~L.~Kane, hep-ph/0202185, and references therein.
    
\bibitem{fcnc} F.~Borzumati and A.~Masiero, 
Phys.~Rev.~Lett. {\bf 57,}  961 (1986);
F.~Gabbiani, E.~Gabrielli, A.~Masiero and L.~Silvestrini,
Nucl.~Phys. {\bf B477,}  321 (1996);
R.~Barbieri, L.~Hall and A.~Strumia,
Nucl.~Phys. {\bf B445,} 219 (1995);  
A.~Bartl, T.~Gajdosik, E.~Lunghi, A.~Masiero, W.~Porod, H.~Stremnitzer
and O.~Vives, Phys.~Rev. {\bf D64,}  076009 (2001),
and references therein.

\bibitem{gauge}
M.~Dine and A.~E.~Nelson, Phys.~Rev. {\bf D48,} 1277 (1993);
M.~Dine, A.~E.~Nelson and Y.~Shirman, Phys.~Rev. {\bf D51,} 1362 (1995);
For a review, G.~F.~Giudice and R.~Rattazzi, Phys.~Rept. {\bf 322,}
419 (1999).

\bibitem{anomaly}
L.~Randall and R.~Sundrum,
Nucl.~Phys. {\bf B557,} 79 (1999);
G.~F.~Giudice, M.~A.~Luty, H.~Murayama and R.~Rattazi,
JHEP {\bf 12,} 027 (1998).

\bibitem{gaugino}
D.~E.~Kaplan, G.~D.~Kribs and M.~Schmaltz,
Phys.~Rev. {\bf D62,}  035010 (2000);
Z.~Chacko, M.~A.~Luty, A.~E.~Nelson and E.~Ponton,
JHEP {\bf 01,} 003 (2000).

\bibitem{string}
A.~Anisimov, M.~Dine, M.~Graesser and S.~Thomas,
hep-th/0111235; hep-th/0201256.
    
\bibitem{ns}
A.~E.~Nelson and M.~J.~Strassler,
JHEP {\bf 09,} (2000) 030; hep-ph/0006251;
T.~Kobayashi and H.~Terao, Phys.~Rev. {\bf D64,} 075003 (2001).
    
\bibitem{nkt}
T.~Kobayashi, H.~Nakano and H.~Terao, Phys.~Rev.
{\bf D65,}  015006 (2002).
    
\bibitem{ls}
M.~A.~Luty and R.~Sundrum,
Phys.~Rev. {\bf D65,}  066004 (2002);  hep-th/0111231.

\bibitem{karch}
A.~Karch, T.~Kobayashi, J.~Kubo, and G.~Zoupanos,
Phys.~Lett. {\bf B441,} 235 (1998);
M.~Luty and  R.~Rattazzi, JHEP {\bf 11,} 001 (1999).
   


\bibitem{antoniadis1}
I.~Antoniadis, Phys.~Lett. B {\bf 246}, 377 (1990); 
I.~Antoniadis, C.~Mu\~noz and M. Quir\'os, 
Nucl.~Phys. {\bf B397,} 515 (1993).

\bibitem{dienes1}
K.~Dienes, E.~Dudas and T.~Gherghetta,
Nucl.~Phys. {\bf B537,} 47 (1999).

\bibitem{arkani1}
N.~Arkani-Hamed, S.~Dimopoulos and G.~Dvali,
Phys.~Rev. {\bf D59,}  086004 (1999).

\bibitem{veneziano}
T.~R.~Taylor and G.~Veneziano, Phys.~Lett. B {\bf 212,} 47 (1988).
   
\bibitem{kobayashi1}
T.~Kobayashi, J.~Kubo, M.~Mondragon and G.~Zoupanos, 
Nucl.~Phys. {\bf B550,} 99 (1999).

\bibitem{polonsky}
N.~Polonsky and A.~Pomarol,
Phys.~Rev.~Lett. {\bf 73,} 2292 (1994).

\end{references}
\end{document}